\documentclass[twocolumn,epjc3]{svjour3}
\journalname{Eur. Phys. J. A}

\usepackage{bm,latexsym,amssymb,amsmath,mathrsfs}
\usepackage{graphicx}
\usepackage{color}
\usepackage{ulem}
\usepackage{setspace}

\newcommand{\A}{\mathcal{A}}
\newcommand{\gluon}{\mathrm{gluon}}
\newcommand{\quark}{\mathrm{quark}}

\begin{document}

\title{Overview of external electromagnetism and rotation in lattice QCD}

\author{Arata Yamamoto\thanksref{add}}
\institute{Department of Physics, The University of Tokyo, Tokyo 113-0033, Japan\label{add}}
\date{Received: 27 February 2021}

\maketitle

\begin{abstract}
This is an introductory review of lattice QCD with external fields.
The study of external magnetic fields is one of the greatest achievements in modern lattice QCD.
Large-scale simulations and detailed analyses have revealed intriguing properties of QCD in the magnetic fields.
The study of external electric fields is more challenging because of a technical difficulty.
We overview the successes and challenges of the lattice simulations with the electromagnetic fields.
We also introduce a newly developing field, the lattice simulation of rotating QCD matters.
\end{abstract}
\section{Introduction}
\label{sec:1}

Quantum chromodynamics (QCD) is the theory of the strong interaction.
Lattice QCD is the most powerful method for the first-principle study of QCD, in particular, of non-perturbative aspects of QCD.
In its long history, it has given us the beneficial knowledges of various phenomena: quark confinement \cite{Wilson:1974sk,Creutz:1980zw}, low-energy properties of hadrons \cite{Aoki:2019cca}, and many others \cite{Gattringer:2010zz,Rothe:1992nt}.
The application of lattice QCD is not restricted to ``pure'' QCD.
It is applicable to QCD coupled with the fields other than quarks and gluons.

Even though the electromagnetic coupling constant $e$ is much smaller than the strong coupling constant $g$, electromagnetism can be relevant for QCD physics if the field strength is so strong that $\sqrt{eE},\sqrt{eB} \sim \Lambda_{\rm QCD}$.
Such intense electromagnetic fields appear in the early Universe, in compact stars, and in heavy-ion collision experiments.
Theoretical physicists predict that the electromagnetic fields induce fascinating phenomena; the magnetic catalysis, the chiral magnetic effect, etc \cite{Miransky:2015ava}.
Are they measurable in the real world?
Are they hidden by any contamination?
Or do they really exist? 
It is not easy to answer these questions.
In the physically interesting cases mentioned above, typical energy scales, the strength of the electromagnetic field, pion mass, and temperature, are comparable.
All of them are unignorable.
We need quantitative calculation based on the microscopic theory.
Lattice QCD would be the best approach for this mission.

A magnetic field is closely related to rotation.
The magnetic field induces the circular motions of charged particles, and vise versa.
Indeed, the intense magnetic field and fast rotation appear in the same places, e.g., in rotating compact stars and in peripheral heavy-ion collisions.
This motivates the study of rotational phenomena in QCD.
The similarity between the magnetic field and rotation is ubiquitous in physics.
For instance, they generate quantum vortices in superconductors or in superfluids \cite{Eto:2013hoa}.
They induce anomalous currents; the magnetic field induces the chiral magnetic effect and the rotation induces the chiral vortical effect \cite{Kharzeev:2015znc}.
One might expect that the same theoretical approach is possible.
From the practical point of view, however, they are quite different.
The difference is essential to see why the rotation is difficult in lattice QCD.
Although the lattice simulation with the rotation is at an early stage of development, it is expected to help us investigate rotating QCD matters.

In this paper, we review the recent activities of lattice QCD with external electromagnetic fields and rotation.
The targets of this review are not only lattice QCD researchers but also non-experts of lattice QCD.
We would like to focus on the formulation and its problem, and to survey what has been done in this field, but not to go into the details of individual simulations.
The readers are encouraged to visit each literature for the detail.

\section{Basics of lattice QCD}
\label{sec:2}

Let us start with the basics of lattice QCD.
In lattice field theories, all the equations in a continuous space are replaced by the counterparts on a discretized space.
The notation is different from the continuous one.
In this review, we write the equations in the continuous notation.
Although this might be unconventional, it would be better for the readers who are unfamiliar with lattice QCD.

Euclidean QCD is formulated in the four-dimensional spacetime $x^\mu=(x,y,z,\tau)$.
The path integral is given by
\begin{equation}
 Z = \int \mathcal{D}\bar{\psi} \mathcal{D}\psi \mathcal{D}A \, e^{-S_\gluon[A]-S_\quark[\bar{\psi},\psi,A]}
\end{equation}
with the gluon action
\begin{equation}
 S_\gluon = \int d^4x \frac{1}{2} {\rm tr} F^{\mu\nu} F_{\mu\nu}
\end{equation}
and the quark action
\begin{equation}
 S_\quark = \int d^4x \bar{\psi} D \psi = \int d^4x \bar{\psi} \left[ \gamma^\mu D_\mu +m \right] \psi ,
\end{equation}
where $D_\mu$ is the SU(3) covariant derivative.
After the Grassmannian integration, we get
\begin{equation}
 Z = \int \mathcal{D}A \, \det D[A] e^{-S_\gluon[A]} .
\label{eqZ1}
\end{equation}
When the theory couples to external fields, the actions are modified.
When external electromagnetic fields exist, the quark action is modified.
The gluon action is not modified because gluons are charge neutral.
The fermion determinant depends on the Abelian gauge field $\A_\mu$,
\begin{equation}
 Z = \int \mathcal{D}A \, \det D[A,\A] e^{-S_\gluon[A]} .
\label{eqZ2}
\end{equation}
When external gravitational fields exist, the actions depend on the metric tensor $g_{\mu\nu}$.
Since gravity couples to all particles, both of the gluon and quark actions are modified as
\begin{equation}
 Z = \int \mathcal{D}A \, \det D[A,g] e^{-S_\gluon[A,g]} .
\label{eqZ3}
\end{equation}
Since these external fields are classical backgrounds, they do not change the integral measure.

In numerical simulations, the functional integral is evaluated by the Monte Carlo sampling method.
The gauge configuration $\{A^a_\mu\}$ is generated according to the probability distribution
$P \propto \det D \ e^{-S_\gluon}$,
and then the statistical average is taken.
The probability distribution must be semi-positive definite.
When the semi-positivity is lost, the Monte Carlo method fails.
This is called the sign problem.
The semi-positivity is satisfied in the original from \eqref{eqZ1}.
The proof is easy.
The gluon action is real and the fermion determinant is given by
\begin{equation}
\begin{split}
 \det D[A] &= \prod_{n} (i\lambda_n+m) 
\\
&= \prod_{\substack{n\\ {\rm s.t.} \lambda_n=0}} m \prod_{\substack{n\\ {\rm s.t.} \lambda_n>0}} (\lambda_n^2+m^2) >0
,
\end{split}
\end{equation}
where the Dirac eigenvalue $\lambda_n$ is defined by $\gamma^\mu D_\mu\phi_n = i\lambda_n \phi_n$.
Here we used the fact that the complex conjugate pair $\pm i \lambda_n$ exist for the nonzero Dirac eigenvalues.
The pairwise property holds only when the massless Dirac operator is anti-Hermitian.
The semi-positivity is not necessarily satisfied when the actions are modified.
A famous example is a quark chemical potential.
For the nonzero quark chemical potential $\mu$, the Dirac operator becomes
\begin{equation}
\begin{split}
 D[A,\mu] &= \gamma^k D_k + \gamma^4(D_4+\mu) + m
\\
&= D[A,\mu=0] + \mu\gamma_4.
\end{split}
\label{eqDmu}
\end{equation}
The additional term breaks the anti-Hermiticity, and thus makes the fermion determinant complex.
Some classes of the external fields are known to cause the sign problem.
The sign problem is a major obstacle in lattice QCD with the external fields.
(Strictly speaking, some lattice fermions, e.g., the Wilson fermion, are not anti-Hermite.
The single-flavor Wilson fermion has the sign problem, but this sign problem can be easily avoided by considering two flavors.
The following arguments are correct even for such lattice fermions.)

\section{Magnetic field}
\label{sec:3}

QCD in external magnetic fields has been intensively studied in the lattice simulations.
One reason is, of course, the physical interest.
Another reason is the technical simplicity.
The Dirac operator is given by
\begin{equation}
\begin{split}
 D[A,\A] &= \gamma^k (D_k+iq\A_k) + \gamma^4 D_4 + m ,
\end{split}
\end{equation}
where $q$ is the electric charge of a quark.
The fermion determinant is semi-positive definite for any choice of $\A_k$.
The most commonly-used setup is the uniform magnetic field along one direction, e.g., $\A_2=Bx$ and $\A_1=\A_3=0$.
(It is possible even on the lattice with periodic boundary conditions \cite{AlHashimi:2008hr}).
The uniform magnetic field has homogeneous effects on physical observables, so greatly simplifies theoretical analysis.

\subsection{Hadron property}

The calculation of hadron masses is a strong area of lattice QCD.
The mass shifts in the magnetic fields were investigated for many hadrons: pseudo-scalar mesons \cite{Bali:2011qj,Hidaka:2012mz,Luschevskaya:2012xd,Luschevskaya:2014lga,Luschevskaya:2015cko,Bali:2017ian,Bignell:2020dze,Ding:2020hxw}, vector mesons \cite{Hidaka:2012mz,Luschevskaya:2012xd,Luschevskaya:2014lga,Bali:2017ian,Lee:2008qf,Luschevskaya:2018chr}, and nucleons \cite{Primer:2013pva,Beane:2014ora,Chang:2015qxa}.
The calculation is straightforward in most cases.
Exceptions are the hadrons with broken quantum numbers.
For example, the uniform magnetic fields break the three-dimensional spherical symmetry SO(3) down to the two-dimensional rotational symmetry SO(2).
The total angular momentum $J$ is not conserved while the projected angular momentum $J_z$ is still conserved.
As a consequence, unpolarized vector mesons ($J=1,J_z=0$) are ill-defined in the magnetic fields.
They are just excited states of pions.

\begin{figure}
\begin{center}
\includegraphics[width=.5\textwidth]{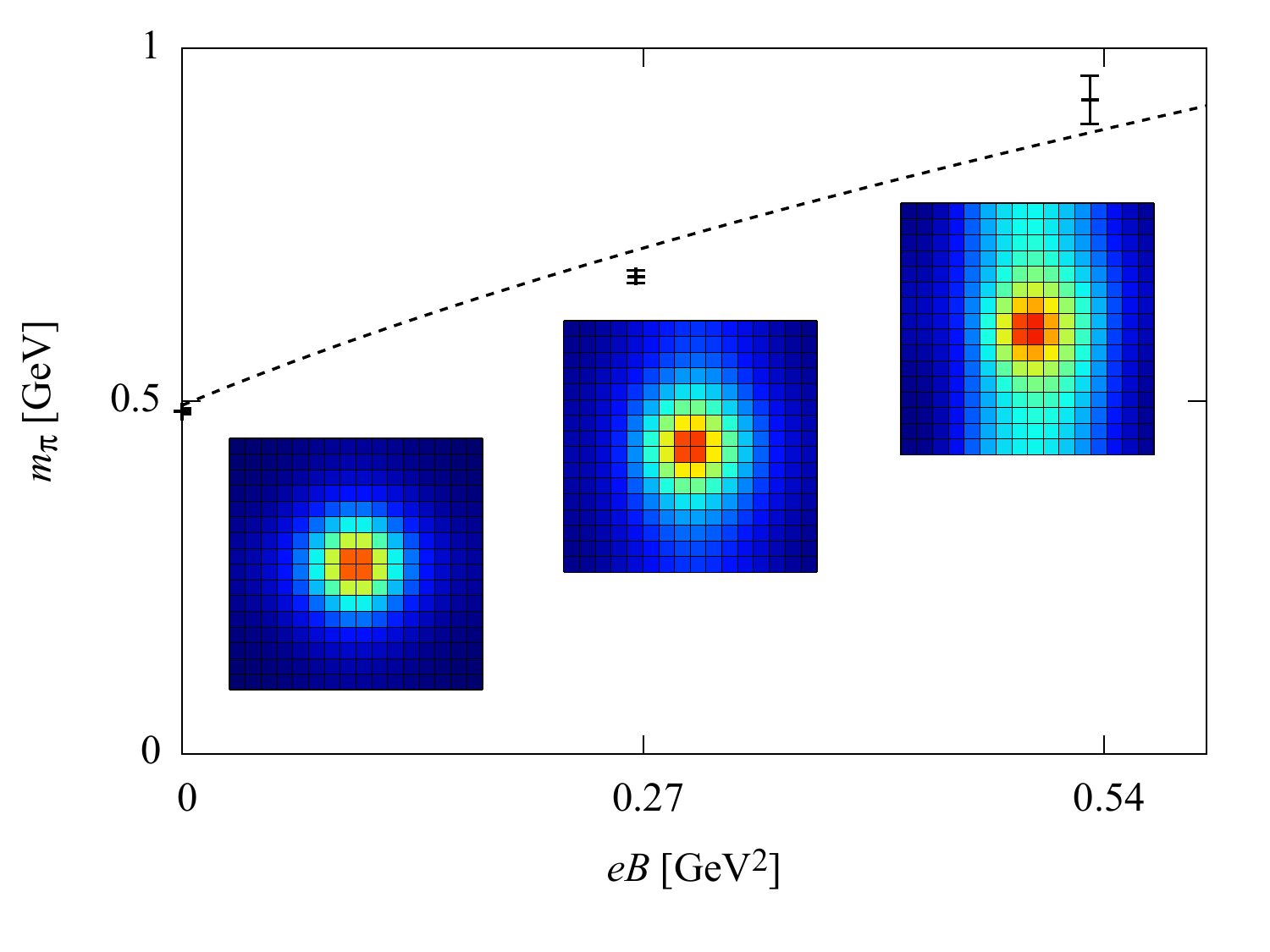}
\caption{Charged pion mass \cite{Hidaka:2012mz} and deformation \cite{Hattori:2019ijy} in external magnetic fields.
The data were obtained by the quenched lattice simulation with unphysical quark mass.
The broken curve is Eq.~\eqref{eqMQM}.}
\label{figM}
\end{center}
\end{figure}

If hadrons were point particles, the masses could be estimated by the Landau quantization in quantum mechanics.
For example, the mass of positively- and negatively-charged pions would be given by
\begin{equation}
 \{ m_\pi(B) \}^2 = \{m_\pi(B=0)\}^2 + |eB|.
\label{eqMQM}
\end{equation}
The magnetic field dependence of the charged pion mass has been investigated in details \cite{Bali:2011qj,Hidaka:2012mz,Luschevskaya:2015cko,Bali:2017ian,Ding:2020hxw}.
One of the results is shown in Fig.~\ref{figM}.
The point particle picture \eqref{eqMQM} is not bad when the magnetic field is weak.
Apart from the weak field limit, the internal structure of hadrons is no longer negligible.
A prominent case is the mass shifts of neutral hadrons.
Since neutral point particles do not interact with the magnetic field, the mass shifts of neutral hadrons come from their compositeness.
The neutral pion mass has small but non-vanishing dependence even if the magnetic field is weak \cite{Hidaka:2012mz,Luschevskaya:2012xd,Luschevskaya:2014lga,Luschevskaya:2015cko,Bali:2017ian,Bignell:2020dze,Ding:2020hxw}.
This indicates the importance of the microscopic calculation based on QCD.
We note that the pion decay constant $f_\pi$ in the magnetic field was also calculated and the Gell-Mann-Oakes-Renner relation \cite{Ding:2020hxw} and leptonic decay rate \cite{Bali:2018sey} were discussed.

The magnetic field breaks spherical symmetry.
The anisotropy can be observed as the shape deformation of charged particles.
The deformation of point particles is predicted by the Landau quantization; the wave function is squeezed in the transverse plane.
Since hadrons are the composites of charged quarks, the deformation is non-trivial.
The shape of a charged pion, which is defined by the distribution of quark number density in the pion \cite{Hattori:2019ijy}, is shown in Fig.~\ref{figM}.
The pion is spinless, so spherically symmetric in the absence of the magnetic field.
When the magnetic field is switched on, the pion is elongated in the longitudinal direction.
The elongation cannot be explained by the Landau quantization, and should be interpreted as a many-body effect.
The anisotropy was also found in the static quark-antiquark potential \cite{Bonati:2014ksa,Bonati:2016kxj,Bonati:2017uvz,Bonati:2018uwh} (and in the gluon correlation function \cite{DElia:2015eey}).
The static quark-antiquark potential is a pure gluonic observable but can be affected by the magnetic field through sea quarks.
The string tension decreases in the longitudinal direction and increases in the transverse direction, which favors the elongated deformation of mesons.

Many lattice studies showed the evidence that the magnetic field plays big roles in hadron physics.
At the same time, they suggested the importance of the ab-initio calculation in lattice QCD.
The magnetic field has impacts on all the microscopic contents, both of the valence and sea quarks, in hadrons.
None of them are negligible.

\subsection{Phase diagram}

QCD matters turn from hadrons to quark-gluon plasmas at a nonzero temperature, although this is not a rigorous phase transition but a crossover (at least, at a zero baryon density).
The QCD phase diagram is usually discussed in terms of spontaneous chiral symmetry breaking and quark confinement \cite{Andersen:2014xxa}.
Let us add one more axis, the axis of magnetic field strength, to the phase diagram.
Does the magnetic field promote or obstruct them?
The order parameter of spontaneous chiral symmetry breaking is the chiral condensate.
The chiral condensate is sensitive to the magnetic field strength.
According to the most renowned scenario, the magnetic catalysis \cite{Gusynin:1994xp}, the magnetic field catalyzes chiral symmetry breaking.
Lattice studies, however, revealed that the chiral condensate in QCD is not so simple \cite{Bali:2011qj,Ding:2020hxw,DElia:2010abb,DElia:2011koc,Bali:2012zg,Bali:2013esa,Bruckmann:2013oba,Ilgenfritz:2013ara,Bornyakov:2013eya,Bali:2014kia,Cea:2015ifa,DElia:2018xwo,Endrodi:2019zrl,Braguta:2019yci,Ding:2020inp}.
The chiral condensate is not a monotonic function of the magnetic field strength; it increases in some parameter regions but decreases in other parameter regions. 
The reason for the non-monotonicity is the interplay of the magnetic catalysis and the ``inverse'' magnetic catalysis \cite{Bruckmann:2013oba}.
The order parameter of the confinement is the Polyakov loop.
The magnetic field dependence of the Polyakov loop is non-trivial.
Since the Polyakov loop is a pure gluonic observable, it indirectly couples to the magnetic field through quark loop diagrams.
The present lattice data suggest that the pseudo-critical temperatures of the Polyakov loop and the chiral condensate move together \cite{Bali:2011qj,DElia:2010abb,DElia:2018xwo,Ding:2020inp}.
The emergence of the first- or second-order phase transition was not found \cite{Endrodi:2015oba}.
Although the phase transition might emerge in an extremely strong magnetic field, the simulation is difficult due to the technical limitation that the magnetic field strength must be smaller than the lattice cutoff scale.

There is another observable to characterize the QCD vacuum in the magnetic field.
The response to the magnetic field is parametrized by the magnetization and magnetic susceptibility, the first and second derivatives of the free energy with respect to the magnetic field strength, respectively.
They are sensitive to a phase of matter.
The quark-gluon plasma phase is paramagnetic, where nonzero magnetization is induced along the direction of the magnetic field, while the hadron phase has very weak magnetic response \cite{Bali:2013esa,Bali:2014kia,Bonati:2013lca,Levkova:2013qda,Bonati:2013vba,Bali:2013owa,Bali:2020bcn}.

\subsection{Chiral anomaly and topological phenomena}

The chiral magnetic effect is the anomalous current generation by the magnetic field in chirally imbalanced matters \cite{Kharzeev:2007jp,Fukushima:2008xe}.
There are two approaches to study the chiral magnetic effect in lattice QCD:
The first approach is to use the chiral imbalance induced by the topological charge of gluons. 
In the QCD vacuum, the total topological charge is zero, $\langle Q \rangle =0$, because of the CP symmetry.
Thus, the one-point function of the vector current is trivially zero, $\langle \bar{\psi} \gamma^k \psi \rangle=0$.
The chiral magnetic effect is measured as the anisotropy of the current fluctuation $\langle (\bar{\psi} \gamma^k \psi)^2 \rangle$ induced by nonzero topological fluctuation $\langle Q^2 \rangle \neq 0$.
This is nothing but the original proposal in heavy-ion collisions \cite{Kharzeev:2007jp}.
The anisotropic current fluctuation \cite{Buividovich:2009zzb,Buividovich:2009wi} and the correlation between electric and topological charges \cite{Abramczyk:2009gb,Bali:2014vja} were investigated in lattice simulations.
The calculation of electric conductivity also supports the chiral magnetic effect \cite{Astrakhantsev:2019zkr} (see also Ref.~\cite{Buividovich:2010tn}).
The second approach is to use the chiral imbalance induced by a chiral chemical potential.
The chiral chemical potential $\mu_5$ is the parameter explicitly breaking the chirality balance, and enables us to observe nonzero vector current, $\langle \bar{\psi} \gamma^k \psi \rangle \propto \mu_5 qB_k$, of the chiral magnetic effect \cite{Fukushima:2008xe}.
The chiral magnetic effect was demonstrated in the lattice simulations with the chiral chemical potential \cite{Yamamoto:2011gk,Yamamoto:2011ks}.
A closely-related effect, the chiral separation effect, was also obtained \cite{Puhr:2016kzp,Buividovich:2020gnl}.

The magnetic fields also play important roles in dense quark matters.
The lattice simulation of dense quark matters is, however, hampered by the sign problem.
(There is an interesting proposal for the sign-problem-free QCD-like theory with nonzero densities and magnetic fields \cite{Brauner:2019rjg}.)
In the dense quark matters, such as the core of neutron stars, topological vortices are created by strong magnetic fields and rotation.
While the topological vortices are universal in physics, a special type of the topological vortices, the ``non-Abelian'' vortices, appear in the high density limit of QCD \cite{Balachandran:2005ev}.
Unfortunately, the sign problem is inevitable in QCD and even in QCD-like theories accompanied by the non-Abelian vortices.
The non-Abelian vortices have been simulated only in a bosonic effective theory, the lattice non-Abelian Higgs model, coupled with the magnetic field \cite{Yamamoto:2018vgg}.

\section{Electric field}
\label{sec:4}

In the Euclidean path integral formalism, there is a definite difference between external magnetic fields and external electric fields.
Let us consider the uniform electric field $E$ along the $z$ direction.
The Abelian gauge field is given by $\A_0=-Ez$ or $E=-\partial_3 \A_0$ in the axial gauge $\A_3=0$.
The Dirac operator is 
\begin{equation}
\begin{split}
 D[A,\A] &= \gamma^k D_k + \gamma^4(D_4-q\A_0) + m
\\
&= D[A,\A=0] -q\A_0\gamma_4.
\end{split}
\end{equation}
This form is the same as Eq.~\eqref{eqDmu}, so shares the same sign problem as the quark chemical potential $\mu$.
The Abelian gauge field $\A_0$ is mathematically equivalent to the chemical potential, except for coordinate dependence.
(The axial gauge choice is just for simplicity. The sign problem cannot be eliminated by an Abelian gauge transformation \cite{Yamamoto:2012bd}.)
The external electric field is very difficult in lattice QCD.

\subsection{Weak electric field}

One naive way to avoid the sign problem is the quenched or partially quenched approximation, where the effects of the electric fields on quark loops are neglected \cite{Shintani:2006xr,Shintani:2008nt}.
The approximation would be valid to some extent in the limit of weak electric field.

In the weak field limit, there is an established scheme applicable to full (i.e., unquenched) QCD.
If the Abelian gauge field is imaginarized as $\A_0 \to \A_4\equiv i\A_0$, the Dirac operator with $\A_4$ is free from the sign problem.
Of course, the imaginarized theory is not the real world.
After performing the simulation with $\A_4$, one must extrapolate the results from the region of $\A_4^2>0$ to the region of $\A_0^2=-\A_4^2>0$.
The extrapolation is justified for small $\A_0$ as long as the theory has analyticity.
The electric polarizability of hadrons \cite{Fiebig:1988en,Christensen:2004ca,Engelhardt:2007ub,Detmold:2009dx,Detmold:2010ts} and the topological $\theta$-term \cite{DElia:2012ifm} were studied in this scheme.
(If you are familiar with the sign problem of dense QCD, you can easily understand this scheme on the analogy of the imaginary chemical potential $\mu_I \equiv i\mu$.)

\subsection{Strong electric field}

A strong electric field is more interesting because it totally changes QCD physics; e.g., the perturbative vacuum is broken down by the Schwinger mechanism \cite{PhysRev.82.664}.
For the strong electric field, there is no general method to avoid the sign problem.
Only one special solution is known.
In two-flavor QCD, the sign problem can be avoided if electric charges are taken as
\begin{equation}
 q_3 \equiv \left( q_u, q_d \right) = \left( +\frac{e}{2}, -\frac{e}{2} \right)
,
\label{eqq3}
\end{equation}
which is called the isospin electric charge \cite{Yamamoto:2012bd}.
The fermion determinant of each flavor is complex, but the phase factors cancel out, and the total fermion determinant is semi-positive.
The lattice simulation with $\A_0$ is possible for any value of $E$.
Although this hypothetical combination of electric charges is different from the physical one, it is useful for the studies of non-perturbative phenomena induced by the strong electric field.
(This is the analogy of the isospin chemical potential $\mu_3 \equiv (\mu,-\mu)$ in dense QCD.)

\begin{figure}
\begin{center}
\includegraphics[width=.5\textwidth]{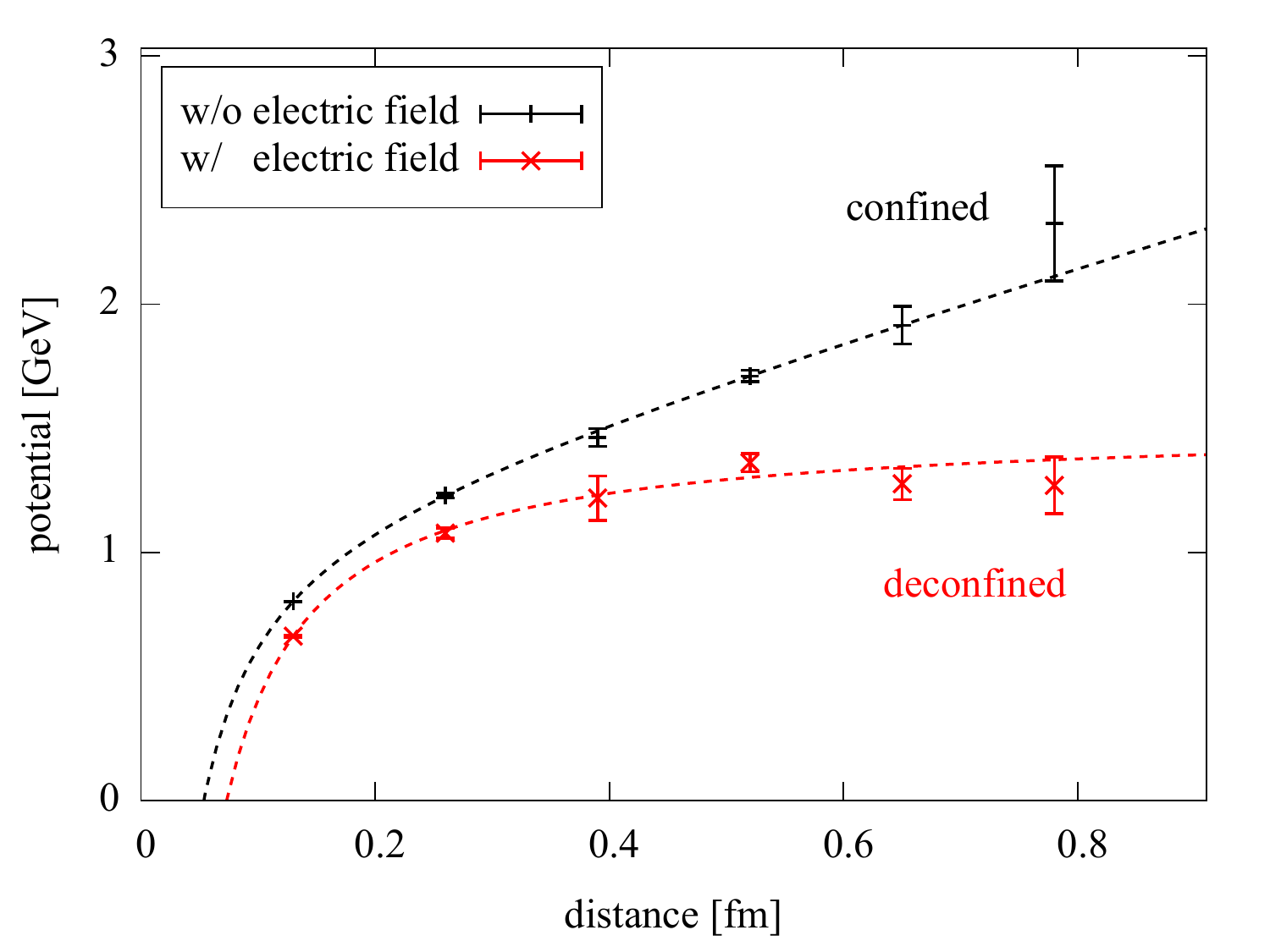}
\caption{Quark-antiquark potential without and with external electric fields \cite{Yamamoto:2012bd}.
}
\label{figWL}
\end{center}
\end{figure}

In Fig.~\ref{figWL}, we show the result obtained by the simulation with the isospin electric charge.
In the QCD vacuum, a quark and an antiquark are bounded by the confining force $\sigma$.
If they are placed in an external electric field, they are dragged by the electric force $qE$ in the opposite direction because they have the opposite electric charges.
When the electric force is stronger than the confining force $qE\ge\sigma$, the confinement is completely lost.
The quark-antiquark potential becomes flat at long distance.
Although the simulation was done with the isospin electric charge, the mechanism is universal for real quark-antiquark pairs.

\section{Rotation}
\label{sec:5}

Numerical simulation of a rotating matter is not easy in general.
Many constituents of the matter are simultaneously moving.
Such collective motion is a highly excited state in the language of quantum theory.
A clever way is to consider the coordinate transformation to a rotating frame.
The rotating matter and its constituents are at rest in the rotating frame, so the simulation is much easier.
As we learned in classical mechanics, the transformation is simple in non-relativity.
The action is shifted as $S(\Omega) = S(\Omega=0) - \Omega L$, where $\Omega$ is the angular velocity and $L$ is the angular momentum of particles.
The ground state has nonzero angular momentum $\langle L \rangle \neq 0$.
In relativistic theories, the rotating frame is formulated as a curved spacetime described by a metric tensor.
We need the formulation of external gravitational fields.

\subsection{Lattice QCD in rotating frames}

In curved spacetimes, the gluon action is
\begin{equation}
 S_\gluon = \int d^4x \sqrt{\det g} \frac{1}{2} g^{\mu\nu} g^{\rho\sigma} {\rm tr} F_{\mu\rho} F_{\nu\sigma}
\label{eqSgrot}
\end{equation}
and the quark action is
\begin{equation}
 S_\quark = \int d^4x \sqrt{\det g} \bar{\psi} \left[ \gamma^\mu(D_\mu+i\Gamma_\mu) +m \right] \psi
\label{eqSqrot}
\end{equation}
with the gravitational connection $\Gamma_\mu$ \cite{Parker:2009uva}.
In the rotating frame around the $z$-axis, the metric tensor is given by
\begin{equation}
g_{\mu\nu}=
{\arraycolsep=3\arraycolsep
\renewcommand\arraystretch{1.5}
 \begin{pmatrix}
  1 & 0 & 0 & -iy\Omega \\
  0 & 1 & 0 & ix\Omega \\
  0 & 0 & 1 & 0 \\
  -iy\Omega & ix\Omega & 0 & 1-r^2\Omega^2
 \end{pmatrix}
},
\label{eqgmunu}
\end{equation}
where $r = \sqrt{x^2+y^2}$ is the distance from the rotation axis.
The lattice theory was formulated by discretizing these actions and metric tensor \cite{Yamamoto:2013zwa}.
Note that lattice field theory in curved spacetimes has been considered in various contexts \cite{Jersak:1996mn,Campos:1998jp,Hayakawa:2006vd,Trencseni:2012su,Yamamoto:2014vda,Villegas:2014dqa,Benic:2016kdk}.
This is one specific application of them.

The actions \eqref{eqSgrot} and \eqref{eqSqrot} with the metric tensor \eqref{eqgmunu} are complex.
Since both of the quark and gluon actions are complex, the sign problem is more severe than that of the chemical potential or the electric field.
In addition, there is another difficulty.
As for the electromagnetic fields, homogeneous field configurations are possible.
As long as the classical theory is translationally invariant, the translational symmetry is preserved even after quantization.
A renormalization constant is independent of coordinate.
On the other hand, the rotation is always inhomogeneous.
The schematic figure is shown in Fig.~\ref{figCVE}.
The rotation axis breaks translational symmetry, which  is seen as the coordinate-dependence of the metric tensor \eqref{eqgmunu}.
Boundaries also breaks homogeneity.
The boundaries must exist in the rotating frame; infinite volume is impossible because the rotational speed $v\equiv \sqrt{x^2+y^2}\Omega$ must not exceed the speed of light.
Once the translational symmetry is violated, the renormalization constant is no longer independent of coordinate.
The artificial inhomogeneity of the renormalization constant must be subtracted from simulation data \cite{Benic:2016kdk}.

Because of these problems, the applicability is now limited to slow rotation.
The present simulations rely on the formulation of the Euclidean rotation \cite{Yamamoto:2013zwa}.
The angular velocity in Eq.~\eqref{eqgmunu} is replaced as $\Omega \to i\Omega_E$ to avoid the sign problem.
One first perform the simulation with the Euclidean rotation, and then perform the analytic continuation from the Euclidean rotation $\Omega^2=-\Omega^2_E<0$ to the Minkowski rotation $\Omega^2=-\Omega^2_E>0$.
The rotational effect on the confinement-deconfinement phase transition was analyzed in this method \cite{Braguta:2020biu,Braguta:2021jgn}.

\subsection{Torsion}

\begin{figure}
\begin{center}
\includegraphics[width=.48\textwidth]{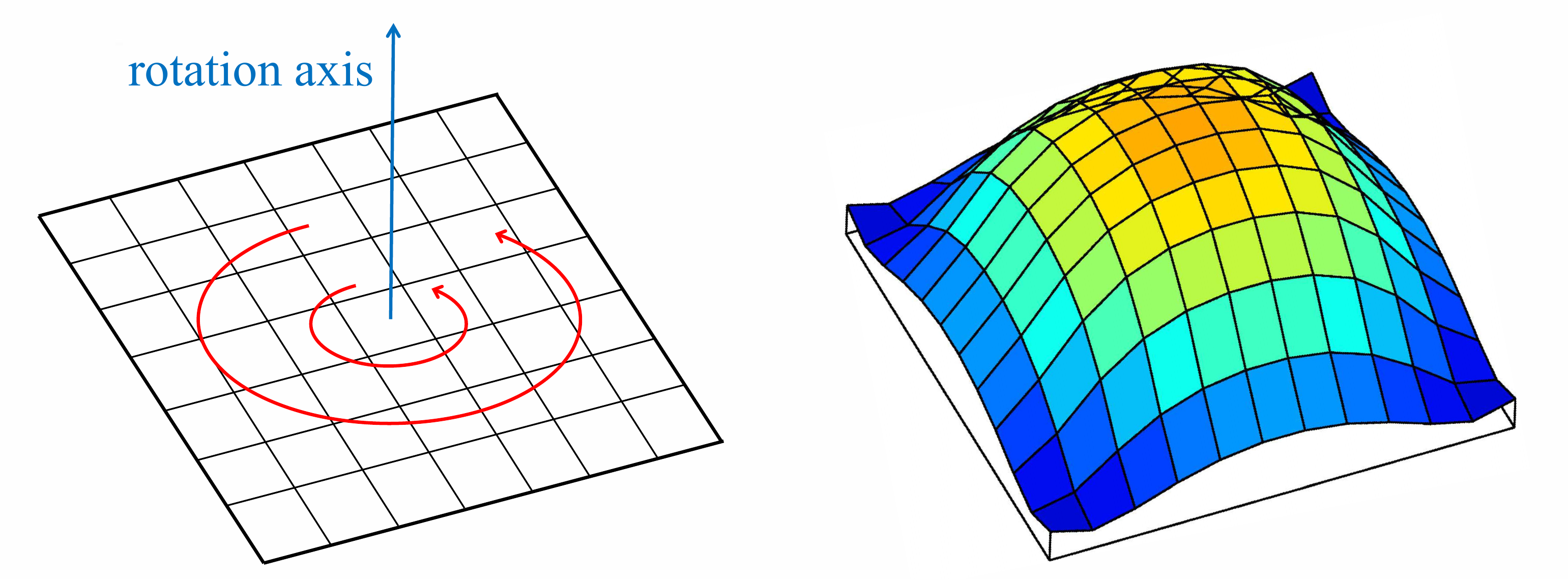}
\caption{Rotating lattice (left) and the chiral vortical effect (right).
The color plot is the strength of the vector current of a free Wilson fermion with open boundary conditions.
}
\label{figCVE}
\end{center}
\begin{center}
\includegraphics[width=.48\textwidth]{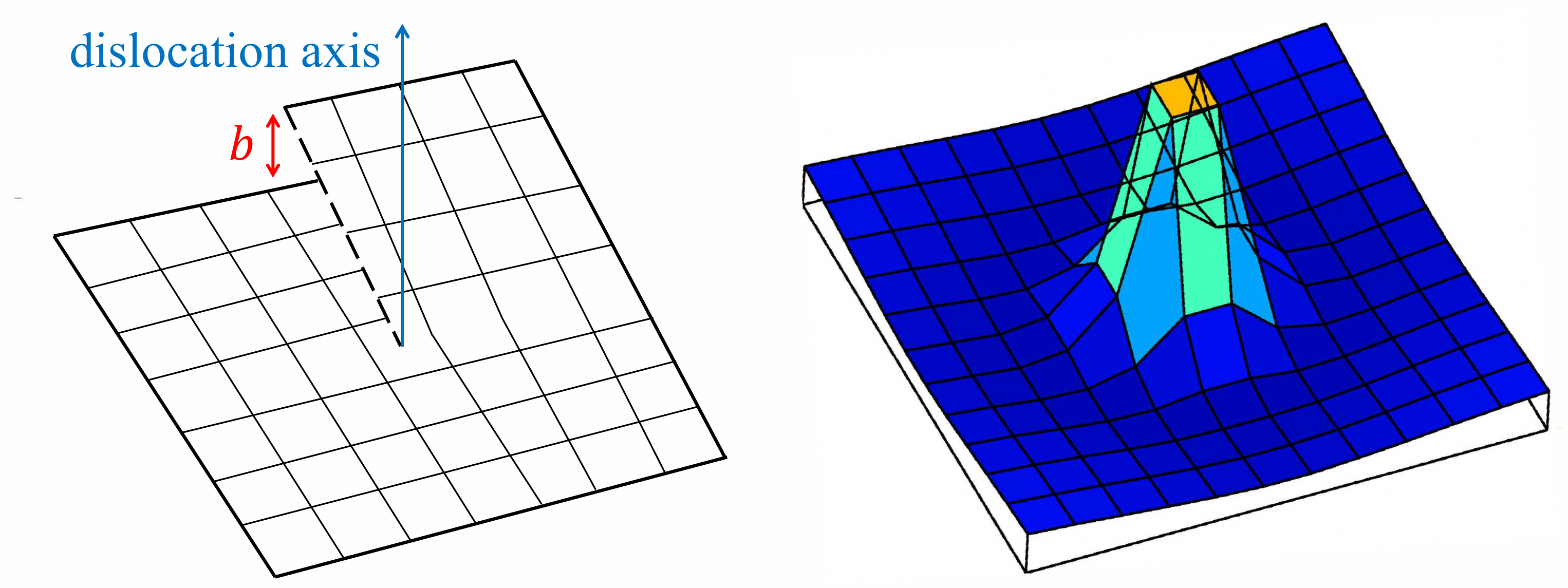}
\caption{Lattice with a screw dislocation (left) and the chiral torsional effect (right) \cite{Imaki:2019ite}.
The color plot is the strength of the vector current of a free Wilson fermion with open boundary conditions.
}
\label{figCTE}
\end{center}
\end{figure}

In the above formulation, the curved spacetime is realized by the metric tensor on the lattice.
There is an alternative lattice formulation of the curved spacetime, that is, changing lattice geometry from a hypercube to something else.
In the second formulation, we can realize the gravitational effects that cannot be realized by the metric tensor.
An example is torsion.
In the spacetime with nonzero torsion, derivatives do not commute, $[\partial_\mu,\partial_\nu]x^\alpha \neq 0$.
The torsion is assumed to be zero in the Einstein gravity, but it is nonzero in a generalized framework of gravity theory, the so-called Einstein-Cartan gravity.
The torsion can be mimicked in laboratory experiments of helical crystals \cite{Kleinert:1989}.
The spacetime with the torsion can be simulated by the lattice with a dislocation \cite{Imaki:2019ite}.
The dislocation is a line defect to distort the lattice.
One type of the dislocations, a screw dislocation, is drawn in Fig.~\ref{figCTE}.
The lattice is spiral around the dislocation axis.

The rotation and torsion induce similar phenomena.
The rotation induces the chiral vortical effect \cite{Kharzeev:2007tn,Son:2009tf} and the torsion induces the chiral torsional effect \cite{Imaki:2019ite,Khaidukov:2018oat,Imaki:2020csc}.
They are shown in Figs.~\ref{figCVE} and \ref{figCTE}.
The chiral vortical effect in Fig.~\ref{figCVE} is obtained with the lattice fermion with the metric tensor \eqref{eqgmunu}.
The constant angular velocity $\Omega$ induces the current
\begin{equation}
 \langle \bar{\psi} \gamma^3 \psi \rangle \propto \mu_5 \mu \Omega
\end{equation}
 in the whole space, although the current is inhomogeneous due to the rotation axis and boundary condition.
The chiral torsional effect in Fig.~\ref{figCTE} is obtained with the lattice fermion with one screw dislocation.
The screw dislocation corresponds to the delta-function torsion in the continuum limit.
The current is induced as
\begin{equation}
 \langle \bar{\psi} \gamma^3 \psi \rangle \propto \mu_5 \mu b \delta(x)
\end{equation}
only at the dislocation axis.
The strength of the torsion is controlled by the parameter $b$, the length of the Burgers vector, which is defined by how many lattice sites are dislocated.

\section{Summary}
\label{sec:6}

The external electromagnetism is one of the active areas of lattice QCD in recent years.
The hadron properties and the QCD phase diagram in the magnetic fields have been investigated at the quantitative level.
It turned out that the magnetic field dependence is not simple.
We need further systematic analysis at the physical point to conclude their fates in the real world.
The electric fields have being less investigated because of the sign problem, but the exceptional solution was found.
The further development is expected.
The lattice study of the rotation has just started.
The rotational effect on QCD matters is a controversial issue even at the qualitative level.
There is plenty of room for lattice QCD to contribute to such a discussion.

Some of these external fields lead non-equilibrium processes; e.g., the electric field induces current flow.
Although non-equilibrium systems are of great interest, the application of lattice QCD is currently restricted to equilibrium systems.
If the restriction is removed in future, real-time dynamics induced by the external fields might become an active area in the next generation of lattice QCD.

\begin{acknowledgement}
The author was supported by JSPS KAKENHI Grant No.~19K03841.
\end{acknowledgement}

\bibliographystyle{epj}
\bibliography{paper}

\end{document}